\documentclass{article}
\usepackage{graphicx}

\begin{document}

\title{Lees-Edwards boundary conditions for Lattice Boltzmann}

\author{ Alexander J. Wagner\thanks{awagner@ph.ed.ac.uk}\\
Department of Physics and Astronomy, University of Edinburgh,\\
                            JCMB Kings Buildings, Mayfield Road,
                            Edinburgh EH9 3JZ, U.K. 
\and Ignacio Pagonabarraga\\
Departament de F\'{\i}sica Fonamental, Universitat de Barcelona, \\
Av. Diagonal 647, 08028-Barcelona, Spain
}
\maketitle

\begin{abstract}
Lees-Edwards boundary conditions (LEbc) for Molecular Dynamics
simulations \cite{LE} are an extension of the well known periodic
boundary conditions and allow the simulation of bulk systems in a
simple shear flow.  We show how the idea of LEbc can be implemented in
lattice Boltzmann simulations and how LEbc can be used to overcome the
problem of a maximum shear rate that is limited to less then $1/L_y$
(with $L_y$ the transverse system size)
in traditional lattice Boltzmann implementations of shear flow.  The
only previous Lattice Boltzmann implementation of LEbc \cite{PRE}
requires a specific fourth order equilibrium distribution. In this
paper we show how LEbc can be implemented with the usual quadratic
equilibrium distributions.
\end{abstract}

\section{Introduction}
In 1972 Lees and Edwards published a seminal paper \cite{LE} describing an
extension of the periodic boundary conditions for Molecular Dynamics
that allows the simulation of a bulk system in a simple shear
flow. The uniform shear flow steady state reached is computationally
convenient because it reduces finite size effects. If moving solid
walls are instead introduced to produce a shear flow, the spatial
inhomogeneities induced close to the wall limit the spatial region
within the simulated system that can be used to study the bulk
behavior of a system under shear. LEbc ensure that the system is
spatially homogeneous, so that bulk-like behavior is recovered on
smaller simulated systems. For lattice Boltzmann methods there is the
additional problem that the maximum shear rate is limited by the
existence of a maximum shear rate of order $1/L_y$ with $L_y$ the
transverse system size. This is required to ensure that all fluid
velocities remain small compared to unity in lattice
units\cite{thesis}.

The original LEbc were implemented for a system of particles, with
positions $(r_x,r_y,r_z)$ and velocities $(v_x,v_y,v_z)$ within a box
of dimension $(L_x,L_y,L_z)$. Lees and Edwards showed that by
implementing a new kind of boundary condition in a MD simulation a
state with a uniform shear velocity profile ${\bf u}=\dot{\gamma} y
{\bf e}_x$ (where ${\bf e}_x$ is the unity vector in x-direction)
could be achieved. The method consists of a simple recipe that imposes
periodic boundary conditions on particles leaving the simulation box
in the directions perpendicular to the velocity gradient ${\bf
e}_y$. Particles leaving the box in the ${\bf e}_y$ direction
($r_y>L_y$) will be reinserted with their $x$-velocity increased by
the shear velocity $\Delta {\bf u}=\dot{\gamma}L_y {\bf e}_x$ added at
a position given by their periodic image displaced by the
time-dependent offset $d_x = t \Delta u_x$, $t$ being the time elapsed
from an appropriate origin of times. If the particles leave the box in
the $-{\bf e}_y$ direction ($r_y<0$) they reappear at the position of
their periodic image displaced by $-d_x$ and have the shear velocity $
\Delta \bf u$ subtracted from their $x$-velocity.

We can describe the boundary conditions by providing the new positions
$r^\prime$ and velocities $v^\prime$ of the particles after the
particles have been moved by the dynamics of the algorithm
\begin{eqnarray}
r^\prime_x &=& \left\{
\begin{array}{lr}
(r_x \bmod L_x) +d_x  & r_z\geq L_z\\
(r_x \bmod L_x)       & 0 \leq r_z < L_z\\
(r_x \bmod L_x) - d_x & r_z < 0
\end{array}
\right.\nonumber\\
r^\prime_y &=& (r_y \bmod L_y)\\
r^\prime_z &=& (r_z \bmod L_z)\nonumber\\
v^\prime_x &=& \left\{
\begin{array}{lr}
v_x +s_x  & r_z\geq L_z\\
v_x       & 0 \leq r_z < L_z\\
v_x - s_x & r_z < 0
\end{array}
\right.\nonumber\\
v^\prime_y &=& v_y \\
v^\prime_z &=& v_z\nonumber
\end{eqnarray}
This is the best known way of implementing boundary conditions that
allow the simulation of a bulk material under shear.

This method works well for all particle-based simulation methods. The
aim of this paper is to adapt this method for lattice Boltzmann
models, where the fundamental quantity is a distribution function,
rather than a set of particles, which is defined on a unit lattice.
In this case, the generalization of LEbc is not straightforward since
the displacement $d_x$ will in general not correspond to a full
integer multiple of the lattice-spacing and it is not clear how to
implement the required velocity shift in a method with a fixed set of
velocities. In an earlier paper \cite{PRE} one of the authors
introduced an interpolation method to deal with the first problem and
a partial Galilean transformation which was used to add momentum. This
approach required the use of an extended equilibrium distribution
function. Hence, it was not applicable to standard lattice-Boltzmann
models.

In this paper we develop an alternative approach to impose
Lees-Edwards boundary conditions in standard lattice-Boltzmann
models. It is based on applying a Galilean-transformation and imposing
the appropriate change of momentum across the moving
plane. Additionally, we must displace the densities appropriately with
an interpolation scheme.  The derived method of Galilean
transformations is quite general and we we can re-derive the correct
boundary conditions for a moving solid wall in contact with the fluid
which were originally introduced by A. Ladd \cite{ladd}.

After describing the lattice-Boltzmann model, we will discuss in
detail the implementation of Lees-Edwards boundary conditions, and we
will subsequently validate the method by examining implementations for
binary fluid mixtures in two dimensions.

\section{Lattice Boltzmann model}
The lattice Boltzmann method can be viewed as a simple discretization
of the Boltzmann equation where space is represented by a lattice and
the one-particle distribution functions (which we will refer to as
`{\em densities}') $f_i({\bf x},t)$ exist at each node ${\bf x}$ of
the underlying lattice, for only a small number of velocities, $\{{\bf
v}_i\}$, that (with unit time step) usually connect a small subset of
lattice points.  These densities evolve following a discrete-time
dynamics. In each iteration, the densities are moved along their
corresponding velocity vectors during the streaming step to the
appropriate neighboring site, while in the subsequent collision step
the set of densities at a given node are relaxed toward
equilibrium. This relaxation step is performed ensuring mass and
momentum conservation at each node. The energy is usually not
conserved (although energy conserving schemes also exist
\cite{teixeira}) but it is possible to introduce a nominal
temperature, that can be viewed as the effect of an underlying
thermostat. This temperature is introduced as a parameter of the
equilibrium distribution function (see below).  The evolution of the
densities in each time step can be written as
\begin{equation}
f_i({\bf x+v}_i,t+1)=f_i({\bf x},t)+\frac{1}{\tau}\left(f^0_i(\rho({\bf
x},t),{\bf u}({\bf x},t))-f_i({\bf x},t)\right)\label{LB}
\end{equation}
where $\tau$ is a relaxation parameter, which in the hydrodynamic
limit of this equation is related to the viscosity of the fluid.
Eq.(\ref{LB}) can be viewed as the discretized version of BGK
approximation of the Boltzmann equation. The linearized collision
operator can be generalized to allow for several relaxation parameters
(and hence the existence of different viscosities for the
fluid)\cite{ladd}. In numerical applications the above evolution
equation is usually split into two subroutines, a collision and a
streaming step. The collision step is local
\begin{equation}
\hat{f}_i ({\bf x},t) = f_i({\bf x},t) + \frac{1}{\tau}\left(f^0_i(\rho({\bf
x},t),{\bf u}({\bf x},t))-f_i({\bf x},t)\right)
\end{equation}
and the streaming step moves densities along their associated velocity
vectors
\begin{equation}
f_i({\bf x+v}_i,t+1)=\hat{f}_i (x,t).
\label{ss}
\end{equation}  
In the above equations, $f_i^0$ is the equilibrium
distribution function that determines the thermodynamics of the fluid,
while $\rho$ and $\rho {\bf u}$ are the local mass and momentum
densities. These are the macroscopic quantities of interest, and can
be obtained from the densities $f_i$ as appropriate moments
\begin{equation}
\rho = \sum_i f_i \;\;\; \rho {\bf u} = \sum_i f_i {\bf v}_i
\end{equation}
The choice of the equilibrium densities is one of the key ingredients of 
the model. They are chosen to recover the equilibrium density and momentum
\begin{equation}
\rho  = \sum_i f^0_i \;\;\; \rho {\bf u} = \sum_i f^0_i {\bf v}_i
\label{f01}
\end{equation}
and the appropriate pressure tensor in equilibrium
\begin{equation}
{\cal P}_{\alpha \beta} = \sum_i f^0_i
(v_{i\alpha}-u_\alpha)(v_{i\beta}-u_\beta)
\label{f02}
\end{equation}
where $\cal P$ is the sum of the thermodynamic pressure tensor $P$ and
an additional term that is required for Galilean invariance of the
system \cite{holdych}:
\begin{equation}
{\cal P}_{\alpha\beta} = P_{\alpha\beta} + \frac{\tau - 0.5}{3}
(u_\alpha \partial_\beta \rho+u_\beta \partial_\alpha \rho)
\end{equation}
Using the thermodynamic pressure tensor $P$ is one of the standard
ways to impose the desired thermodynamic behavior of the
lattice-Boltzmann fluid\cite{orlandini}.  In the simplest case, the
pressure tensor for the ideal gas is $P_{\alpha\beta}=\rho \frac{1}{3}
\delta_{\alpha\beta}$. We need an additional constraint for the third
moment
\begin{equation}
\sum_i f^0_i v_{i\alpha} v_{i\beta} v_{i\gamma} =
\frac{\rho}{3}(u_\alpha \delta_{\beta\gamma}+ u_\beta
\delta_{\alpha\gamma} + u_\gamma \delta_{\alpha\beta})
\end{equation}
to ensure that the anisotropy of the underlying lattice does not affect 
the effective behavior of the model in the hydrodynamic limit. It is possible 
to fulfill these constraints by introducing equilibrium densities that depend 
quadratically on the lattice velocities, $\{{\bf v}_i\}$. Depending on 
the geometry of the lattice, there is still some freedom in the choice 
of the equilibrium densities.

A Taylor expansion to second order in the derivatives\cite{thesis} of
the zeroth and first moments in ${\bf v}$ of the lattice Boltzmann
equation (\ref{LB}) gives the macroscopic equations which express mass
and momentum conservation. From (\ref{LB}) we have 
\begin{equation}
\sum_{n=1}^\infty \frac{1}{n!}(\partial_t + {\bf v}_i \nabla)^n f_i = 
\frac{1}{\tau} (f^0_i-f_i)
\end{equation}
and we can approximate 
\begin{equation}
f_i = f^0_i + \tau (\partial_t f^0_i + v_{i\alpha}\partial_\alpha
f^0_i) + O(\partial^2).\label{approx}
\end{equation}
The zeroth velocity moment of (\ref{LB}) up to $O(\partial^2)$ gives
the continuity equation
\begin{equation}
\partial_t \rho + \partial_\alpha (\rho u_\alpha) = 0
\end{equation}
and the first velocity moment gives the Navier Stokes equations
(with $\partial_\alpha {\bf u u u} = O(\partial^3)$)
\begin{equation}
\rho \partial_t u_\alpha + \rho u_\beta \partial_\beta u_\alpha =
-\partial_\beta P_{\alpha\beta} + \partial_\beta \left[\rho
\nu  \left(\partial_\beta u_\alpha+\partial_\alpha u_\beta
-\frac{2}{D} \partial_\gamma u_\gamma
\delta_{\alpha\beta}\right)\right]
\label{NS}
\end{equation}
where the viscosity is $\nu =(\tau-0.5)/3$ and $D$ is the number of
spatial dimensions. In order to simulate a mixture of two components
$A$ and $B$ we follow Orlandini\cite{orlandini} and define a second
lattice Boltzmann equation
\begin{equation}
g_i({\bf x+v}_i,t+1)=g_i(x,t)+\frac{1}{\tau_\phi}\left[g^0_i(\rho({\bf
x},t),\phi({\bf x},t),{\bf u}({\bf x},t))-g_i({\bf x},t)\right]
\label{LB2}
\end{equation}
where the density difference (concentration) $\phi=\rho_A-\rho_B$ is defined as
\begin{equation}
\phi = \sum_i g_i
\label{eq:phi0}
\end{equation}
and is the only additional variable conserved in the dynamics.
The total density is now $\rho=\rho_A+\rho_B$ and the concentration
couples back into the momentum equation, eq.(\ref{NS}), via the pressure
tensor $P_{\alpha\beta}$ which is now a function of both $\rho$ and $\phi$. 
The relaxation step for the dynamics of $g_i$ is dependent on 
$\tau_{\phi}$ which is related to the concentration diffusivity in 
the hydrodynamic limit. The equilibrium densities $g_i$ are determined, 
again by imposing that the local equilibrium concentration is recovered 
(from eq.(\ref{eq:phi0})). The additional constraints on the $g^0_i$ are
\begin{equation}
\phi u_\alpha = \sum_i g^0_i v_{i\alpha}, \;\;\;
{\cal M}_{\alpha\beta}= \sum_i g^0_i (v_{i\alpha}-u_\alpha)
(v_{i\beta}-u_\beta)
\end{equation}
Analogously to the pressure tensor, $\cal M$ is given by a correction term
for Galilean invariance \cite{awagner} and the chemical
potential $\mu$:
\begin{equation}
{\cal M}_{\alpha\beta} = \Gamma \mu \delta_{\alpha\beta} + \frac{\tau_\phi
-0.5}{3} (u_\alpha \partial_\beta \phi + u_\beta \partial_\alpha \phi)
\end{equation}
The desired thermodynamics of the fluid is determined by $\mu$ which
must be consistent with the chosen $P_{\alpha\beta}$. These two
quantities are normally derived from a given free
energy\cite{thesis,orlandini}. In this way one ensures their
consistency, and can fix beforehand the appropriate thermodynamic
behavior of the binary fluid model. It is enough to consider equilibrium
densities $g_i$ that depend quadratically on the velocities to impose
the previous requirements.

Again, we recover the hydrodynamic limit of eq.(\ref{LB2}) performing 
a Taylor expansion of to second order in spatial gradients. In this 
case we get a convection-diffusion equation for the concentration
\begin{equation}
\partial_t \phi + \partial_\alpha (\phi u_\alpha) = \partial_\alpha
 \left\{ (\tau_\phi-0.5)
 \left[ \partial_\beta \left(\Gamma \mu(\rho,\phi) \delta_{\alpha\beta}\right)
 + \frac{\phi}{\rho}
 \partial_\beta P_{\alpha\beta}\right] \right\}
\end{equation}
This model has been used extensively for the simulation of two-phase
flows \cite{PRE,thesis,holdych,orlandini,kaare}.The following
derivation of the LEbc presented in the next section is, however, even
more general and can be applied to most lattice Boltzmann methods.

\section{Lees Edwards' boundary conditions}

The implementation of Lees Edwards boundary conditions for lattice
Boltzmann will follow the same steps as in its original
implementation: we will apply periodic boundary conditions to the
distribution function in the two directions perpendicular to the
velocity gradient; and in the third, we will assume that the periodic
system is moving as a whole with a velocity $\Delta \bf u$ with respect
to the lattice. In a second step, we will have to displace the
densities $f_i$ (and $g_i$ for the binary model), taking into account
that we will have to interpolate the densities from the off-lattice
positions to which they stream, onto the regular lattice. So this
boundary condition will be applied after the collision and before the
streaming of the densities.

Let us first focus on the velocity transformation. Since we are
dealing now with a distribution function, the way to implement the
velocity jump when crossing the system boundary in the velocity
gradient direction will be to perform a Galilean transformation, {\it i.e.}
if the distribution function in the system has momentum $\rho {\bf
u}$, its image node will have momentum $\rho ({\bf u}+\Delta {\bf u})$ if we
move in the direction of the gradient, and
$\rho ({\bf u}-\Delta {\bf u})$ when we leave the system in the direction
opposite to the velocity gradient. In this way, we will generate a
linear velocity profile with shear rate $\dot{\gamma}{ \Delta \bf
u}/L_y$.  In order to get the momentum transfer, we need to calculate
the difference between a density $f_i$ at the system
velocity and the one after the Galilean transformation has been
applied. All the macroscopic variables and their derivatives will be
the same, since the image system is moving as a block at constant
velocity. So, using eq.(\ref{approx}), where we retain only the
$u$-dependence of the $f$'s, we obtain
\begin{eqnarray}
\Delta f_i &=&f_i({\bf u}+\Delta{\bf u})-f_i({\bf u})
\nonumber\\
&\approx&f^0_i({\bf u}+\Delta{\bf u})-f^0_i({\bf u})\nonumber\\&&
-\tau \{\partial_t [f^0_i({\bf u}+\Delta{\bf u})-f^0_i(-{\bf u})]
+v_{i\alpha} \partial_\alpha [f^0_i({\bf u}+\Delta{\bf u})-f^0_i({\bf u})]\}
\end{eqnarray}
where $\Delta f_i$ is the change in the density moving in the $i$
direction due to the Galilean transformation. Note that the
transformation will only be applied to the densities crossing the LE
boundary but not to the rest of the densities at the same node.

The terms linear in $\tau$ are of order $O(\partial)$ and their zeroth
and first moments are $O(\partial^3)$. These terms are therefore
small and enters only into the pressure moments.  Therefore we can
neglect them. We obtain for the boundary condition
\begin{equation}
f_i^\prime = f_{i} +f^0_i({\bf u}+\Delta{\bf u})-f^0_i({\bf u}).
\end{equation}
where $f_i^\prime$ means the density after the application of the
boundary condition. 
We now need to specify the equilibrium distribution. We will assume 
the usual quadratic distribution function which can fulfill the
conditions (\ref{f01}) and (\ref{f02}):
\begin{equation}
f_i^0 = \rho a_0^i + \rho a_1^i v_{i\alpha} u_\alpha
+ a_2^i v_{i\alpha} v_{i\beta} \Pi_{\alpha\beta}
+ a_3^i \mbox{tr}(\Pi) \label{equil}
\end{equation}
where $\Pi_{\alpha\beta} = P^0_{\alpha\beta} + \rho u_\alpha u_\beta$ 
and we obtain
\begin{eqnarray}
f^0_i({\bf u}+\Delta{\bf u})-f^0_i({\bf u})
&=& \rho [a_1^i v_{i\alpha} \Delta u_\alpha
\nonumber \\&&
+ a_2^i v_{i\alpha} v_{i\beta} (u_\alpha \Delta u_\beta+
u_\beta \Delta u_\alpha +\Delta u_\alpha \Delta u_\beta)\nonumber \\&&
+ a_3^i (2{\bf u}.\Delta{\bf u}+|\Delta {\bf u}|^2)] \label{qres}
\end{eqnarray}
If we have a binary mixture, the same transformation will apply to the 
densities $g_i$ related to the concentration. The equilibrium 
distribution $g_i^0$ can be written in the same form as 
eq.(\ref{equil}) if we replace $\rho$ by $\phi$ and
 $P_{\alpha\beta}$ by $\Gamma\mu\delta_{\alpha\beta}$. 
In the previous expression $\Gamma$ is a factor which, together with 
the relaxation parameter $\tau_{\phi}$, will determine the diffusivity 
of the concentration; $\delta_{\alpha\beta}$ is the Kronecker delta function.

Now that we have defined a suitable Galilean transform, we have to
split the densities into an untransformed part and a transformed part
that will stream over the boundary, {\it i.e.} the densities at the
top and bottom boundaries with associated velocities $v_{iy}>0$ and
$v_{iy}<0$ respectively if we take ${\bf e}_y$ as the direction of the
shear gradient. It is important to ensure that the two subsets of
velocities conserve the node density after the Galilean transformation
has been applied. This requires

\begin{eqnarray}
\rho &=& \sum_{v_{iy}> 0} f_i^\prime + \sum_{v_{iy}\le 0}
f_i\nonumber\\
&=& \sum_{v_{iy}>0} \left[f_i + f^0({\bf u+\Delta u})-f^0({\bf u})\right] 
+ \sum_{v_{iy}\le 0} f_i\nonumber\\
&=& \sum_{v_{iy}>0} \left[ f^0({\bf u+\Delta u})-f^0({\bf u})\right]
+ \sum_i f_i\nonumber\\
\Leftrightarrow 0 &=& \sum_i \frac{(v_{iy}^2 + v_{iy})}{2}
 \left[f^0({\bf u+\Delta
u}) - f^0({\bf u})\right]\nonumber\\
&=&\frac{1}{2}  (P_{yy} + \rho u_y u_y - P_{yy} - \rho u_y u_y)
\label{eq:consist}
\end{eqnarray}
where $\sum_{v_{iy}>(<) 0}$ is performed over the velocity directions
with a positive (negative) component crossing the shear
boundary. Eq.(\ref{eq:consist}) is fulfilled if we require that the
velocity subset that characterizes the lattice Boltzmann model
involves only velocities that do not span beyond the closest site
layer, {\it i.e.}  the factor $(v_{iy}^2 + v_{iy})/2$ is 1 for
velocities with $v_y>0$ and $0$ otherwise. This is the case for most
standard velocity sets and for the velocity set we will consider here
(see eq.(\ref{v2}))
 in particular. To derive this result we have also used that the
velocity-shift is parallel to the boundary, {\it i.e.} $\Delta u_y=0$.

It is worth remembering that the total momentum in the direction of
the velocity gradient in the system must be zero; otherwise there will
be a net mass transfer across the LE boundaries which would result in
an increase (or decrease) of momentum in the direction of the velocity
gradient. The mean velocity will then steadily increase until it
exceeds the maxium velocity and the simulation becomes numerically
unstable.  This is a general requirement of any Lees-Edwards boundary
condition, because the presence of LE boundaries breaks the momentum
conservation unless the mass transfer is exactly balanced in both
directions.

The idea of the Galilean transformation that we have derived also has
more general applications and can be used to derive the boundary
conditions for a solid moving wall.  At a solid wall, the velocity of
the fluid must be that of the wall. This no-slip boundary condition
for a stationary wall is implemented through the `bounce-back' method
\cite{frisch}. In it, one replaces the streaming step of eqn.(\ref{ss})
for the densities that would have been streamed through the wall with
\begin{equation}
f_i({\bf x},t+1) = f_{-i}({\bf x},t)
\end{equation}
where $v_{-i} = -v_i$. To generalize this bounce back mechanism for a
moving wall we first perform a Galilean transformation of the
densities into a reference frame in which the wall is at rest. We then
perform the bounce back and then transform the densities back with an
inverse Galilean transformation. We then obtain for the bounce back
off a wall moving with velocity $U$
\begin{equation}
f_i({\bf x},t+1) = f_{-i}({\bf x},t) + 2 \rho a_1^2 v_{i\alpha}
U_\alpha
\end{equation}
which is a result first derived with a different method by
Ladd\cite[eq.(3.3)]{ladd}. For this particular case, the change 
in the distribution function is linear in the velocity jump, 
rather than quadratic, as in the general case (cf. eq.\ref{qres}) 
\footnote{
According to the discussion of the previous paragraph, the density 
at both sides of the boundary is not modified by the bounce in the 
case of a moving parallel solid wall that slides along one of the 
boundaries of the system. For more general situations, this is not 
the case. For example, in the case of a moving rigid object embedded 
in the fluid, its velocity is not parallel to the boundary. In 
this case, the density inside the rigid object will be modified 
as a function of time. However, 
the presence of a rigid interface separating the inner from the 
outer motion of the fluid makes it possible to correct 
for these deviations\cite{heemels}.}.

To complete the implementation of Lees-Edwards boundary conditions
 we also need to move the densities
by the displacement of the image system,
 $d_x=t \Delta u_x$, which is not going to be an integer in
general. We must then split the displacement in an integer part $d_x^I$
and a real part $d_x^R$ with $ 0\leq d_x^R < 1$. We then
shift the densities that will stream over the boundary with $v_{iy}>0$
and $v_{iy}<0$ respectively by
\begin{equation}
f^\prime ({\bf x}^\pm,t) = (1-d_x^R) f({\bf x}+d_x^I,t) +
d_x^R f({\bf x}+d_x^I\pm 1,t)
\label{interpol}
\end{equation}
where $x^\pm$ are the positions of the lattice points directly below
or above the LE boundary and this shift and interpolation is applied
to densities $f_i$ with $v_{iy}=\pm 1$.  That is, we have performed a
linear partitioning of the moving density between the two
corresponding nearest nodes. If necessary, it might be possible to
improve upon this linear interpolation. This simple choice, however,
already gives very accurate results.

Lees Edwards boundary conditions for lattice Boltzmann were first
derived by Wagner \& Yeomans \cite{PRE} whose approach also required
\begin{equation}
\Delta f_i =  f^0_i(-{\bf u})-f^0_i({\bf u})
\end{equation}
but the authors also insisted on a heuristic transformation rule where
the amount of velocity added to the densities that stream over the
boundary should correspond to $\Delta \bf u$. In particular they first
calculated a separate x-velocity for the particles that crossed the
LEbc and then they required that this x-velocity should be independent
of the y-velocity in equilibrium. In order to fulfill this additional
requirement they needed a more complicated equilibrium distribution (a
polynomial fourth order in ${\bf u}$). While this additional
requirement is perfectly sensible in the limit of the continuous
Boltzmann equation our results show that it is not always helpful to
consider the densities to be those of real particles and that it is
perfectly feasible to have a Lees Edwards boundary condition for a
lattice Boltzmann method with a traditional quadratic equilibrium
distribution.

{
\renewcommand{\arraystretch}{1.8}
\begin{table}
\begin{center}
\begin{tabular}{|l|c|c|c|c|}
\hline
& $a_0^i$ & $a_1^i$ & $a_2^i$ & $a_3^i$\\
\hline
$i=0$       & $1 - \frac{4}{3} (1-k)$ & 0   & 0   & $-\frac{l+1}{l+2}$\\
\hline
$i$=1,\dots,4 & $\frac{2}{3} (1-k)$ & $\frac{1}{3}$ & $\frac{1}{2}$ & $-\frac{1}{2} \frac{1}{l+2}$\\
\hline
$i$=5,\dots,8 & $-\frac{1}{3} (1-k)$ & $\frac{1}{12}$ & $\frac{1}{8}$ & $-\frac{1}{8} \frac{l}{l+2}$\\
\hline
\end{tabular}
\end{center}
\caption{ The coefficients for the D2Q9 model. There are two free
parameters that we set to $k=l=1$ for convenience.}\label{table1}
\end{table}
}

\section{Numerical Implementation}
We implemented a lattice with dimensions $L_x, L_y$ 
and the boundaries of the
simulation lattice have periodic boundary conditions.
We implemented a  D2Q9 model, a two dimensional model with velocities 
{\renewcommand{\arraystretch}{0.8}
\setlength{\arraycolsep}{0pt}
\begin{equation}
{\bf v}_i =
\left\{ 
\left(\begin{array}{r}0\\0\end{array}\right),
\left(\begin{array}{r}1\\0\end{array}\right),
\left(\begin{array}{r}-1\\0\end{array}\right),
\left(\begin{array}{r}0\\1\end{array}\right),
\left(\begin{array}{r}0\\-1\end{array}\right),
\left(\begin{array}{r}1\\1\end{array}\right),
\left(\begin{array}{r}-1\\1\end{array}\right),
\left(\begin{array}{r}-1\\-1\end{array}\right),
\left(\begin{array}{r}1\\-1\end{array}\right)
\right\}\label{v2}
\end{equation}
and with the coefficients given by Table 1.

\subsection{Multiple LE planes}
One of the limitations of Lattice-Boltzmann models to study fluids
under shear arises because the maximum velocity cannot be larger than
one lattice spacing per time step (and in practice must be
$\stackrel{<}{\sim} 0.1$). This gives an upper bound of order $1/L_y$
for the shear rates that can be reached with the method described so
far (one cannot simply increase the lattice size while keeping the
shear rate constant). This suggests that large systems with
Lees-Edwards boundary conditions should be restricted to small
shear-rates. We can now overcome this limitation using the present
method by introducing a larger number of Lees-Edwards shear planes as
one increases the $L_y$ dimension of the simulation. We will still
require, of course, that the velocity in each of the sub-lattices
separated by the LEbc are no larger than about 0.1.

We will allow for  $N^{LE}$ Lees-Edwards shear planes at almost equidistant 
positions
\begin{equation}
y = Y^k = \left[(k-1/2)
L_y/N^{LE}\right]_I
\mbox{ for }k=\{1, ... ,N^{LE}\}
\end{equation}
where $[x]_I$ is the largest integer smaller than x. The system is then
effectively cut into $N^{LE}-1$ sub-lattices that will move relative
to each other with the associated jump-velocities $U^k$.  The
average shear rate for the system is therefore
\begin{equation}
\dot{\gamma} = \frac{\sum_{k=1}^{N^{LE}} U_k}{L_y}
\label{gammadot}
\end{equation}
and the stationary solution due to this set of LE shear planes is a linear
shear profile of the form
\begin{equation}
{\bf u} = \left(\begin{array}{c}
		\dot{\gamma} (y-y_0^k)\\ 0\\0
		\end{array}
	  \right)
\end{equation}
where $y_0^k$ is the position of zero velocity for each
sub-lattice. 

At this point it may appear that the introduction of the shear-planes
does introduce a local velocity profile and biases the flow but, in
the absence of errors, the only the global shear rate $\dot{\gamma}$,
defined in eq.(\ref{gammadot}) matters to the flow. This is easily
seen if one considers a Molecular Dynamics simulation where the
introduction of several shear planes is equivalent to describing the
different sublattices in different frames of references, with no
effect on the real dynamics. Because of this there is, of course, no
merit in introducing more than one LE-plane (which establishes
$\dot{\gamma}$) in MD simulations.  The only limit for the use of many
LE-planes in lattice Boltzmann lies in the errors induced by the
additional internal boundary condition and we discuss them in the
following sections.

\begin{figure}
\begin{center}
\includegraphics[width=12cm]{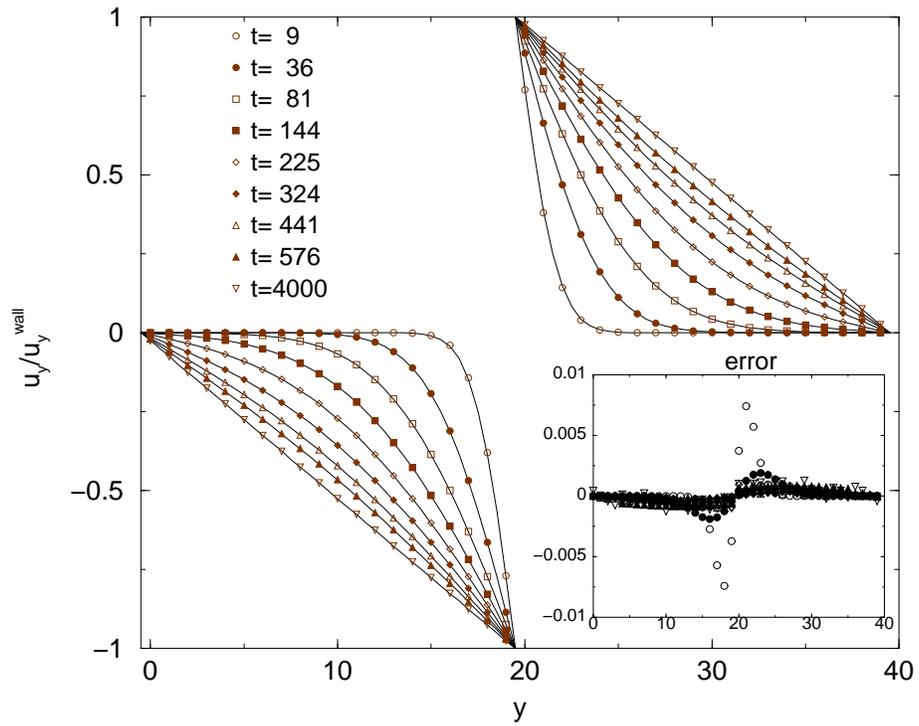}
\end{center}
\caption{Start-up of a shear profile. Solid lines are given by
eq. (\ref{startth}) and the symbols represent simulation data. The
inset shows the difference of the theoretical and simulated velocities.}
\label{fig1}
\end{figure}

\subsection{Transient start of shear and stationary profile} 
As a first test for the new boundary conditions we will examine the
accuracy with which the algorithm simulates the start-up of a shear in
a single component fluid. We consider a system with one LEbc and the
initial velocity is zero. This corresponds to a system in which there
is a periodic stack of slabs of fluid of width $L_y$ each of which
moves with a different velocity. The velocity of the nth slab is given
by $u_x=n \Delta u$. The system will then exhibit viscous friction and
the profile will converge toward a linear shear-profile. The density
and pressure in this system remain constant and the velocity only has
an $x$-component and its value only changes in the $y$-direction. The
Navier-Stokes equation for this system gives
\begin{equation}
\partial_t u_x = \partial_y^2 u_x
\end{equation}
{\it i.e.} a simple diffusion equation for the momentum. The solution
for this equation for $L_y=\infty$ is well know \cite[\S 52]{ll} and
given by
\begin{equation}
u_x(y,t) = \mbox{erf}\left( \frac{y}{2 \sqrt{\nu t}}\right)
\end{equation}
where 
\begin{equation}
\mbox{erf}(x) = \frac{2}{\sqrt{\pi}} \int_0^x e^{\xi^2} d\xi.
\end{equation}
Obtaining the solution for finite $L_y$ is simply a linear
superposition
\begin{equation}
u_x(y,t) = \sum_{k=-\infty}^\infty \frac{k}{|k|} 
\left[1-\mbox{erf}\left(\frac{y+kL_y}{2\sqrt{\nu t}}\right)\right]
\label{startth}\end{equation}
which can be conveniently evaluated numerically. The agreement
between the theoretical prediction and the measured velocity is very
good as can be seen in Figure \ref{fig1}. The error shown in the inset
of Figure \ref{fig1} is largest for the
early time simulations but is always below 1\% of the maximum velocity.

\subsection{Shear in a two component model}
The previous simulation was effectively a one-component simulation and
it did not test the effectiveness of the interpolation
scheme. Unfortunately, it is not easy to find two-component flows under
shear for which analytical solutions are known. There is, however, a
method to test the errors introduced by the LEbc for a two-component
system very accurately. We can choose to have two equidistant LEbc
with equal but opposite velocities so that the effective shear-rate is
zero. The resulting system is then equivalent to a system at rest. We
will consider two periodic stripes of A-rich and B-rich material at
rest and simulate it with two LEbc with opposite velocities.

In figure \ref{figure2} we show the results of a simulation on a 40x60
lattice with a stripe that was 20 lattice spacings wide. There were
two LEbc with velocities of 0.005 and -0.005. The figure shows a section
of 20x20 after 20000 iterations ($d_x=0$) where the stripe interface
overlaps the LE boundary. The full system, of course, comprises of two
interfaces in the x-direction and two LEbc in the y-direction and
there the same error-patterns appear by symmetry.

Without errors this system would show perfectly straight contour lines
and velocities of 0.005 in the upper half of figure \ref{figure2}(a)
and -0.005 in the lower half. Since the velocities are very nearly
that, we have to examine the difference of the real velocity and the
predicted velocity shown in figure \ref{figure2}(b). We see that this
error corresponds to a deviatoric velocity profile that flows into the
interface at the LEbc. We attribute this error to the fact that we use
the linear interpolation from eq.(\ref{interpol}). The effect of
this interpolation is to smear out the interface and make it wider
than the equilibrium profile. There will then be a Marangoni flow that
tries to relax the profile to its equilibrium shape which we assume is
the major contribution to this velocity profile.  A close examination
of the interface reveals a slight widening around the LEbc. The
maximum amplitude of the error in
the velocity is  3\%. In future work we plan to examine if we can
reduce this error by using an interpolation method that will conserve
not only the mass but also the derivatives and prevents the additional
diffusion introduced by the LEbc. Nonetheless, the accuracy found above
is comparable to other sources of error ({\it e.g.} anisotropy)
present in lattice Boltzmann as currently used.

\begin{figure}
\begin{minipage}{6cm}
\begin{center}
\includegraphics[height=5.5cm]{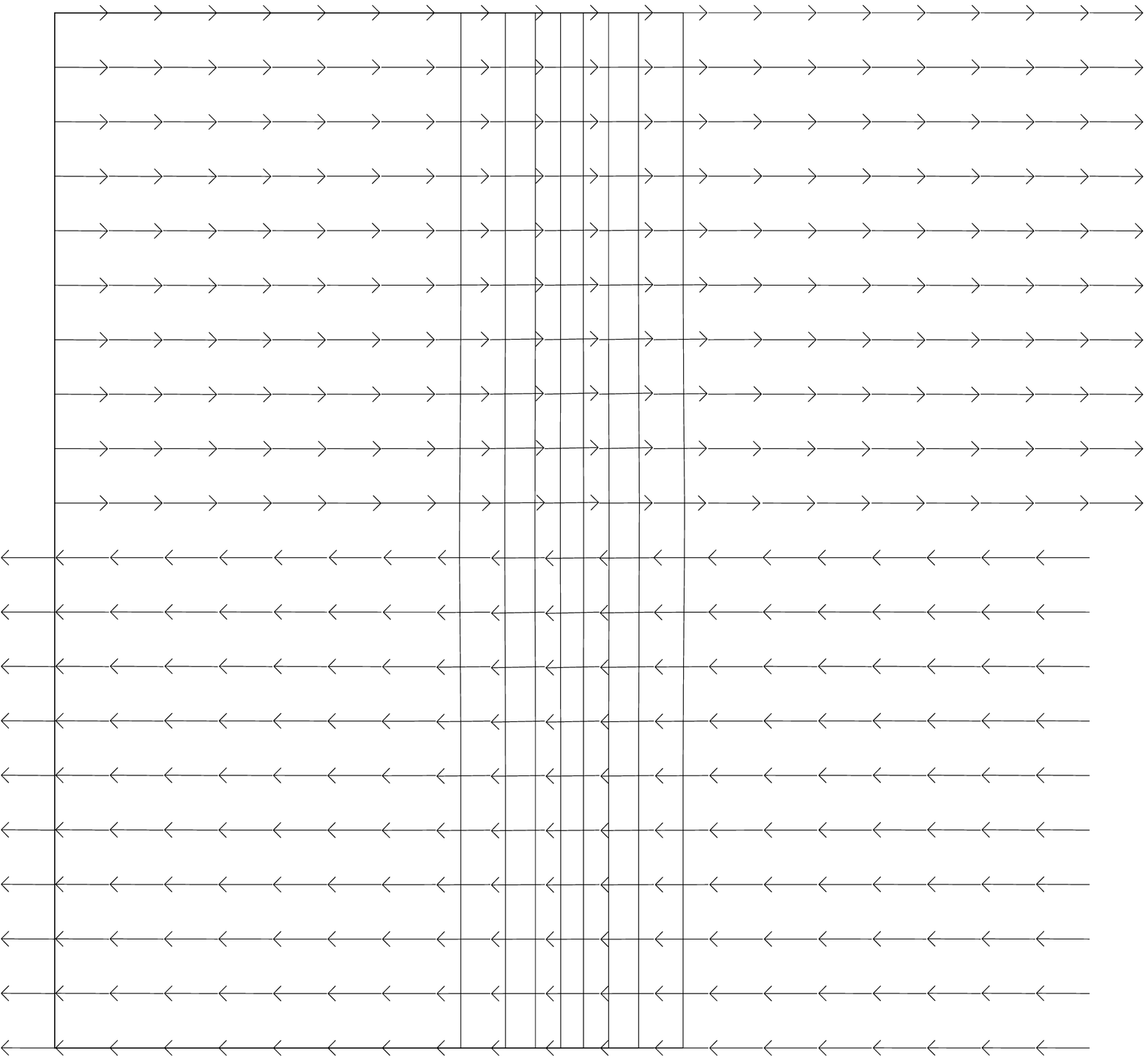}\\
(a)
\end{center}
\end{minipage}
\begin{minipage}{6cm}
\begin{center}
\includegraphics[height=5.5cm]{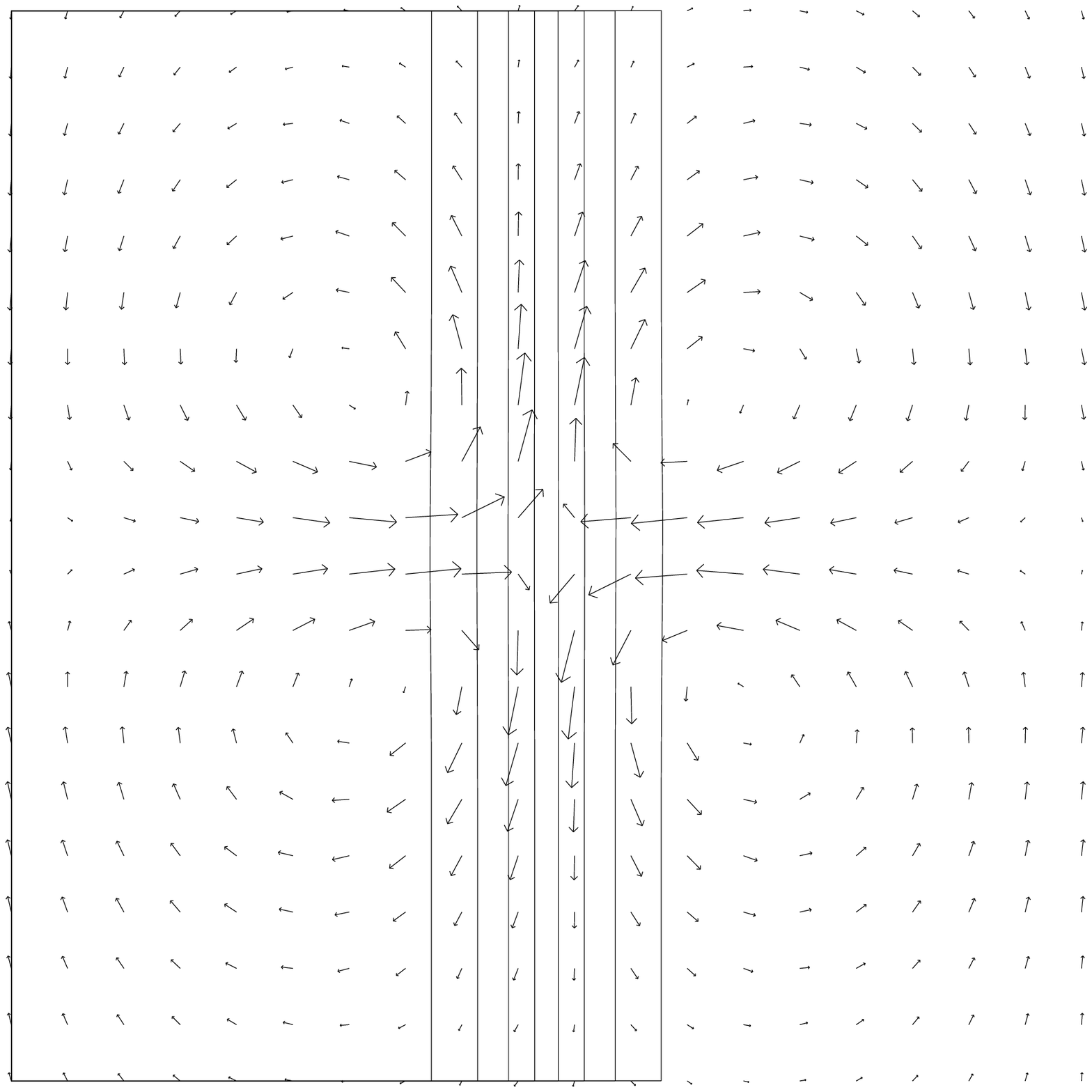}\\
(b)
\end{center}
\end{minipage}
\caption{Steady state profile of the stripe morphology. In (a) the
real velocities are shown and the maximum velocity corresponds to
0.0051. In (b) the deviation from the theoretical profile is shown and
the maximum velocity here is 0.00013. Eight contour lines are also
shown indicating the width of the interface. (See text for details)}
\label{figure2}
\end{figure}

As a final benchmark of the performance of LEbc's, we have studied the
shape of a drop in a uniform shear flow generated by a single LE
plane.  In order to check for artifacts induced by the presence of the
boundary, we have compared a simulation where the drop does not cross
the boundary, and a situation where the LE plane bisects the drop.  In
Fig.\ref{fig:drop} we show the comparison for two different capillary
numbers $Ca$ -defined as the ratio between viscous and interfacial
forces, $Ca=\nu R \dot{\gamma}/ \sigma$, with $R$ the radius of the
drop and $\sigma$ its surface tension. The interface is located in the
simulation through an interpolation of the corresponding nearby nodes
where the order parameter changes sign, which introduces a small
source of error. The drop that is bisected by the boundary will have a
larger averaged velocity. Since we are comparing the drop shapes at
equal time, we have displaced the drop accordingly to make them
overlay. In the figure one can clearly see that the intersection of
the boundary does not affect the shape of the drop, within the
numerical accuracy already mentioned in the previous paragraph.

\begin{figure}
\begin{minipage}{6cm}
\begin{center}
\includegraphics[height=5.5cm]{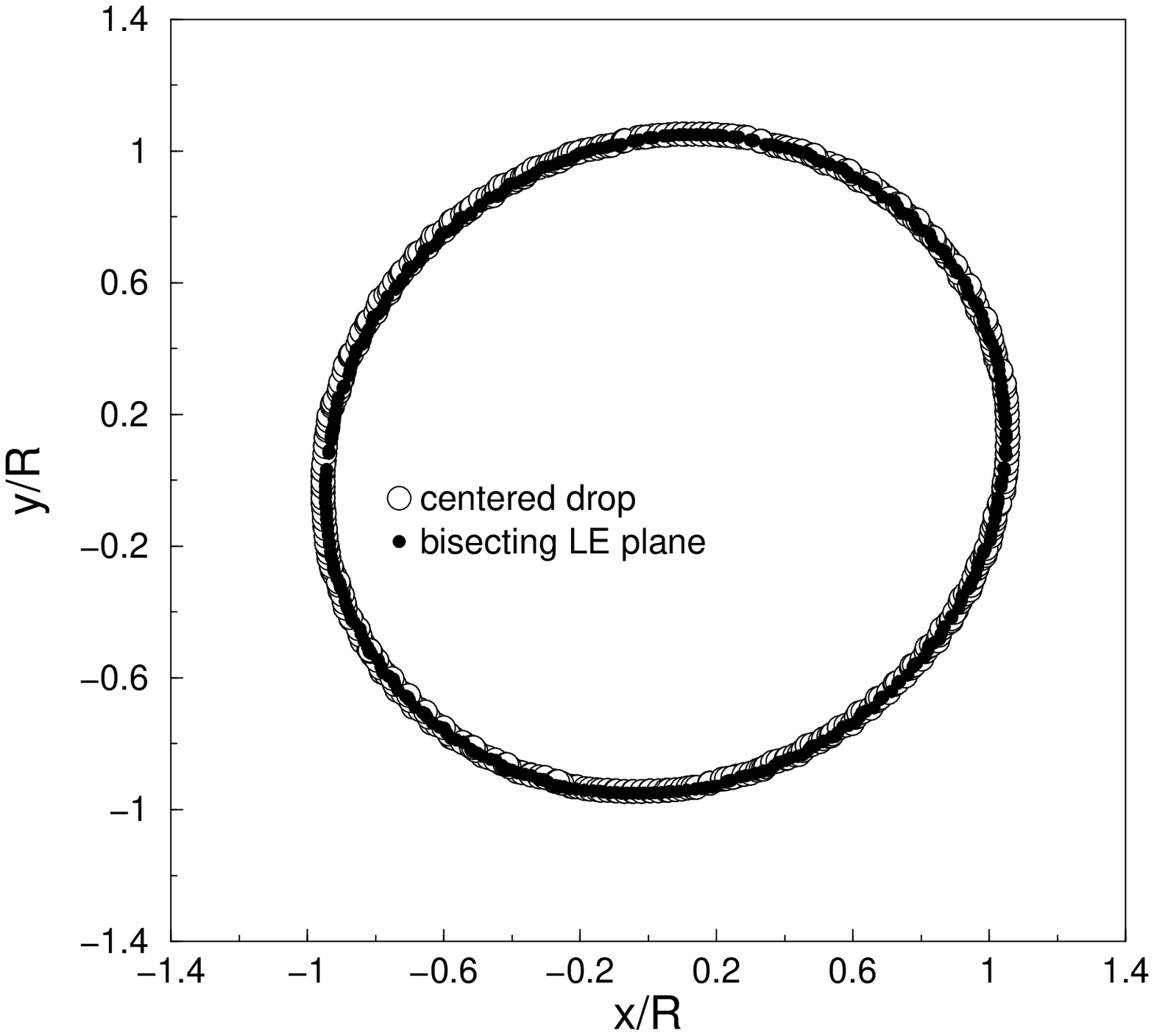}\\
(a)
\end{center}
\end{minipage}
\begin{minipage}{6cm}
\begin{center}
\includegraphics[height=5.5cm]{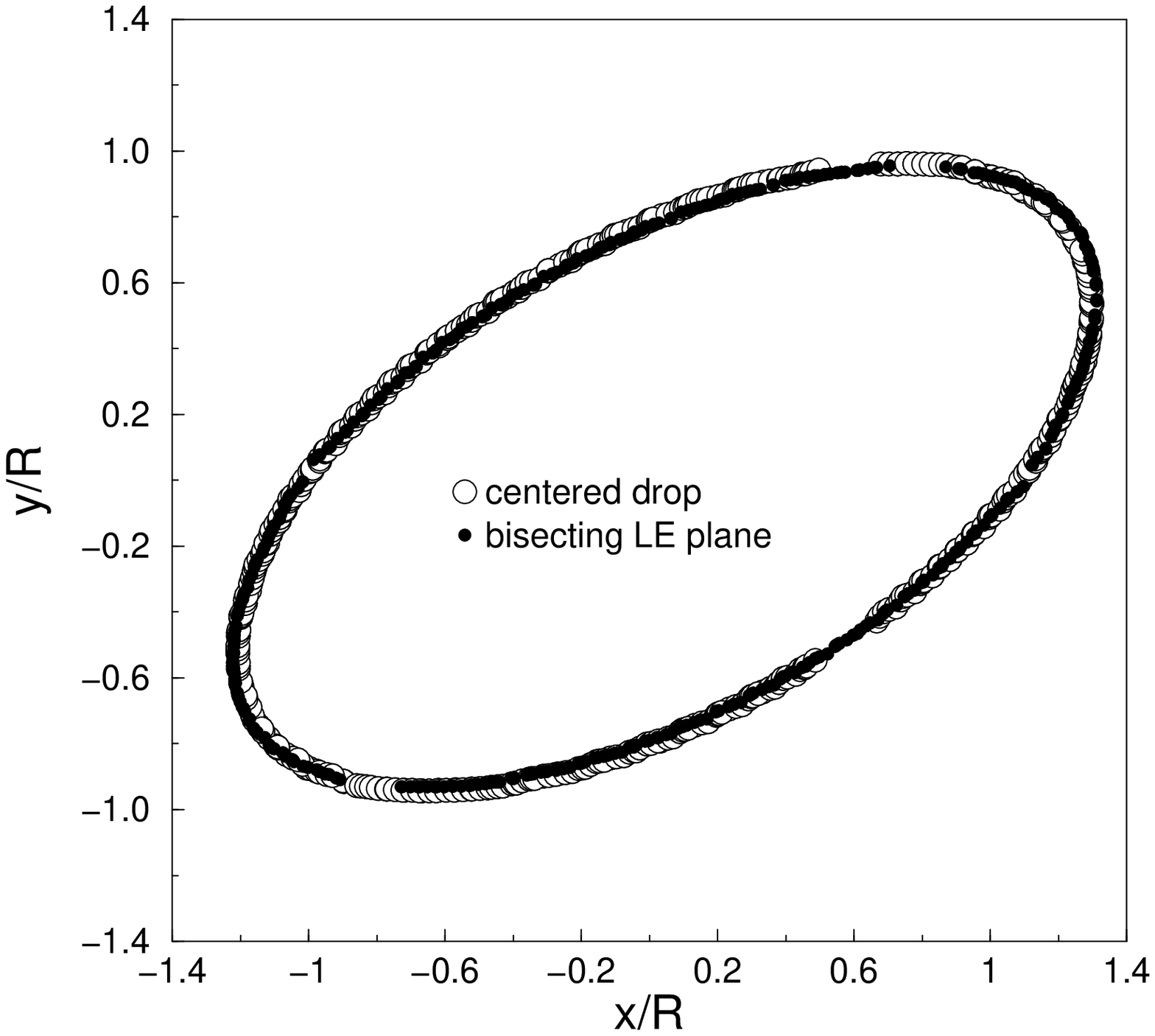}\\
(b)
\end{center}
\end{minipage}
\caption{Steady shapes of a drop in a Couette flow for different capillary 
numbers. In the open circle situation the drop does not intersect the LE plane.
  The filled symbols correspond to the simulation where the LE plane 
bisects the drop. a) $Ca=0.012$, b/ $Ca=0.5$. The interfacial width in 
both cases is of the order of 3 in lattice units.}
\label{fig:drop}
\end{figure}

\section{Conclusions}

In this paper we have described how to implement Lees-Edwards periodic
boundary conditions for lattice-Boltzmann models.  Lees Edwards
boundary conditions are a useful for the simulation of bulk systems
under shear without the complicating influence of walls. In this way,
finite size effects are diminished, which has converted the uniform
shear flow in an attractive steady configuration to analyze
computationally.

The implementation we have described follows the basic steps of the
original paper\cite{LE}, and is compatible with the standard LB
models, that assume an equilibrium distribution for the densities
which is quadratic in the velocity. The numerical examples described
show that the method is quantitatively accurate for single and even
for two component systems.

We have also described how to avoid the limitations on the attainable
shear rates that are imposed by the underlying lattice. By adding a
number of shearing planes, steady states of large shear rates can then
me reached (subject only to the intrinsic stability limitations of any
lattice-Boltzmann schemes). Because of the errors introduced by the
interpolation scheme, it is not advisable to introduce more LEbc's
than necessary. One could think of a system where the whole lattice is
subject to LEbc but this would now lead to a non-isotropic diffusion.
We are planing to improve the interpolation scheme in order to reduce
the artificial diffusion.

\section*{Acknowledgements}
We thank M.E. Cates for his critical reading of this manuscript. The
work was funded in part under EPSRC Grant GR/M56234 and EC Access to
Research Infrastructure Contract HPRI-1999-CT-00026 TRACS programme at EPCC.

\def\jour#1#2#3#4{{#1} {\bf #2}, #3 (#4).}
\def\tbp#1{{\em #1}, to be published.}
\def\tit#1#2#3#4#5{{#1} {\bf #2}, #3 (#4)}
\def\ap{Adv. Phys.}
\def\epl{Euro. Phys. Lett.}
\def\prl{Phys. Rev. Lett.}
\def\pr{Phys. Rev.}
\def\pra{Phys. Rev. A}
\def\prb{Phys. Rev. B}
\def\pre{Phys. Rev. E}
\def\pa{Physica A}
\def\ps{Physica Scripta}
\def\zpb{Z. Phys. B}
\def\jmpc{J. Mod. Phys. C}
\def\jpc{J. Phys. C}
\def\jpcs{J. Phys. Chem. Solids}
\def\jpco{J. Phys. Cond. Mat}
\def\jf{J. Fluids}
\def\jfm{J. Fluid Mech.}
\def\arf{Ann. Rev. Fluid Mech.}
\def\roy{Proc. Roy. Soc.}
\def\rmp{Rev. Mod. Phys.}
\def\jsp{J. Stat. Phys.}
\def\pla{Phys. Lett. A}

\end{document}